\documentclass[11pt]{article}
\pdfoutput=1
\usepackage{dcolumn}
\usepackage{bm}

\usepackage{graphicx}
\usepackage{amssymb,amsmath}
\usepackage{multirow}
\usepackage{cite,color,url}
\usepackage[colorlinks=true
,urlcolor=blue
,anchorcolor=blue
,citecolor=blue
,filecolor=blue
,linkcolor=blue
,menucolor=blue
,linktocpage=true
,pdfproducer=medialab
,pdfa=true
]{hyperref}

\usepackage{slashed}
\usepackage{epsfig,psfrag,rotating,soul}
\usepackage{rotfloat}

\evensidemargin \oddsidemargin
\allowdisplaybreaks

\let\OLDthebibliography\thebibliography
\renewcommand\thebibliography[1]{
  \OLDthebibliography{#1}
  \setlength{\parskip}{0pt}
  \setlength{\itemsep}{0pt plus 0.3ex}
}

\begin{document}
\thispagestyle{empty}

\def\thefootnote{\fnsymbol{footnote}}

\begin{flushright}
IFT-UAM/CSIC-19-78\\
FTUAM-19-12\\
LPT-Orsay-19-14
\end{flushright}

\vspace*{1cm}

\begin{center}

\begin{Large}
\textbf{\textsc{Model-independent search strategy for the\\[0.25em] lepton-flavor-violating heavy Higgs boson decay to $\bm{\tau\mu}$ at the LHC}}
\end{Large}

\vspace{1cm}

{\sc
Ernesto~Arganda$^{1}$%
\footnote{{\tt \href{mailto:ernesto.arganda@fisica.unlp.edu.ar}{ernesto.arganda@fisica.unlp.edu.ar}}}%
, Xabier Marcano$^{2}$%
\footnote{{\tt \href{mailto:xabier.marcano@th.u-psud.fr}{xabier.marcano@th.u-psud.fr}}}%
, Nicol\'as~I.~Mileo$^{1}$%
\footnote{{\tt \href{mailto:mileo@fisica.unlp.edu.ar}{mileo@fisica.unlp.edu.ar}}}%
, Roberto A.~Morales$^{3}$%
\footnote{{\tt \href{robertoa.morales@uam.es}{robertoa.morales@uam.es}}}%
and Alejandro Szynkman$^{1}$%
\footnote{{\tt \href{mailto:szynkman@fisica.unlp.edu.ar}{szynkman@fisica.unlp.edu.ar}}}%
}

\vspace*{.7cm}

{\sl
$^1$IFLP, CONICET - Dpto. de F\'{\i}sica, Universidad Nacional de La Plata, \\ 
C.C. 67, 1900 La Plata, Argentina

\vspace*{0.1cm}

$^2$Laboratoire de Physique Th\'eorique, CNRS, \\
Univ. Paris-Sud, Universit\'e Paris-Saclay, 91405 Orsay, France

\vspace*{0.1cm}

$^3$Departamento de F\'{\i}sica Te\'orica and Instituto de F\'{\i}sica Te\'orica, IFT-UAM/CSIC,\\
Universidad Aut\'onoma de Madrid, Cantoblanco, 28049 Madrid, Spain

}

\end{center}

\vspace{0.1cm}

\begin{abstract}
\noindent
In this work we present a model-independent search strategy at the LHC for heavy Higgs bosons decaying into a tau and a muon, $H/A \rightarrow \tau\mu$, showing a plausible tendency to improve the sensitivity obtained by the present experimental limits. 
This search strategy is performed for the Higgs boson mass range 1-5 TeV and uses as the most relevant kinematical variables, in order to discriminate signal against background, the transverse momenta of the muon and the tau together with the missing transverse energy.
We estimate the exclusion limits at 95\% C.L. and the significances for evidence and discovery at $\sqrt{s}$ = 14 TeV with $\cal{L}$ = 300 fb$^{-1}$, observing a growth in the sensitivities for high Higgs boson masses. 
Moreover, since the Higgs boson decay into a tau-lepton pair may mimic our LFV signal, we also study the impact of the ditau channel on the exclusion limits and the significances for evidence and discovery. In particular,
the impact on the exclusion limits of LFV heavy Higgs boson decays is significant when the ditau rate begins to compete with the corresponding to the $H/A \rightarrow \tau\mu$ decay. 
\end{abstract}

\def\thefootnote{\arabic{footnote}}
\setcounter{page}{0}
\setcounter{footnote}{0}

\newpage

\section{Introduction}
\label{intro}

Since the discovery in 2012 of the Higgs boson at the LHC, reported by the ATLAS~\cite{Aad:2012tfa} and the CMS~\cite{Chatrchyan:2012xdj} collaborations, with a mass $m_h =$ 125.09 $\pm$ 0.21 (stat.) $\pm$ 0.11 (syst.) GeV~\cite{Aad:2015zhl}, an intense experimental program has been developed in order to figure out if there is new physics behind it and, in particular, an extended Higgs sector. 
A clear signal of physics beyond the standard model (BSM) would be undoubtedly the presence of Higgs boson decays into two charged leptons of different flavor. The first search at the LHC of lepton-flavor-violating (LFV) decays of the Higgs boson was performed by CMS~\cite{Khachatryan:2015kon}, observing a slight excess with a 2.4$\sigma$ significance in the $h \to  \tau\mu$ channel at a center-of-mass energy of $\sqrt{s} =$ 8 TeV and an integrated luminosity of 19.7 fb$^{-1}$. Later on, ATLAS found a mild deviation of 1$\sigma$ significance in the same LFV channel~\cite{Aad:2015gha}, corresponding also to $\sqrt{s} =$ 8 TeV with 20.3 fb$^{-1}$ of luminosity. In a subsequent CMS analysis at 13 TeV~\cite{CMS:2016qvi} no excess was observed but more data were needed to make definitive conclusions on the origin of that anomaly. Finally, CMS confirmed the disappearance of this excess with the results presented in~\cite{Sirunyan:2017xzt}. Lepton flavor violation has also been searched for in the $\mu e$~\cite{Khachatryan:2016rke} and $\tau e$~\cite{Khachatryan:2016rke,Aad:2016blu} channels of the Higgs boson. This class of LFV processes is also being sought through the decays of heavy resonances~\cite{Aad:2015pfa,Aaboud:2016hmk,Sirunyan:2018zhy,Aaboud:2018jff} and neutral heavy Higgs bosons~\cite{Aaij:2018mea,CMS:2019tiw}.

LFV was firstly observed in the neutrino oscillations and one could expect that also occurs in the charged lepton sector of the SM. Indeed, LFV is intensely looked for through radiative decays ($\mu \to e \gamma$~\cite{TheMEG:2016wtm}, $\tau \to e \gamma$~\cite{Aubert:2009ag}, and $\tau \to \mu \gamma$~\cite{Aubert:2009ag}), leptonic decays ($\mu \to eee$~\cite{Bellgardt:1987du}, $\tau \to eee$~\cite{Hayasaka:2010np}, $\tau \to \mu\mu\mu$~\cite{Hayasaka:2010np}, etc), and $\mu-e$ conversion in heavy nuclei~\cite{Dohmen:1993mp,Bertl:2006up}. No evidence of LFV has been observed in any of these searches, imposing very restrictive bounds on the rates of these LFV processes. In this work we focus on the search for LFV in the decays of a heavy Higgs scalar $H$ and a heavy pseudoscalar $A$. In principle, one should consider the three possible LFV Higgs boson decay (LFVHD) channels: $H/A \to \mu e$, $H/A \to \tau e$, and $H/A \to \tau\mu$. 
The former represents the cleanest LFV signature at hadron colliders, however the present stringent upper limits on the related LFV lepton processes ($\mu \to e \gamma$, $\mu \to eee$, and $\mu-e$ conversion in heavy nuclei) seem to indicate that $H/A\to \mu e$ would be very suppressed. 
In addition, the LFVHD rates are usually proportional to the masses of the heaviest lepton involved in the decay. Therefore, this would mean also very tiny branching ratios for the $H/A \to \mu e$ channel compared to the LFVHD with $\tau$ leptons. In that sense, we expect very similar rates for the second channel, $H/A \to \tau e$, to the latter one, $H/A \to \tau\mu$. Nevertheless, the $\tau e$ channel leaves a more contaminated signature than the $\tau\mu$ channel, due specially to the fact that the jet fake rates are much larger for electrons than for muons. For all these reasons, we will concentrate along this work only on the most promising LFVHD channel, $H/A \to \tau\mu$, for which there have been proposed several search strategies at the LHC, in a model-independent way~\cite{Harnik:2012pb,Davidson:2012ds,Bressler:2014jta} and within the framework of specific BSM models~\cite{Buschmann:2016uzg,Primulando:2016eod,Hou:2019grj}.

We develop a model-independent search strategy for LFV heavy Higgs boson decays at the LHC, that shows a plausible tendency to improve the sensitivity obtained by the current experimental limits, specially in the region of large Higgs boson masses. Our search strategy is performed in three Higgs boson mass windows ($1-1.5$ TeV, $1.5-2.5$ TeV, $2.5-5$ TeV) under the hypothesis that the decay rate into a tau-lepton pair is negligible.
We focus on this mass range due to the fact that the current experimental searches become weaker with increasing values of the mass since the cuts stop being efficient. Besides, we do not consider mass values larger than 5 TeV because heavier particles are hardly produced with the current LHC energy.
The most relevant kinematical variables in order to discriminate signal against background turn out to be the transverse momenta of the muon ($p_T^\mu$) and the tau ($p_T^\tau$), together with the missing transverse energy ($E_T^\text{miss}$), being the latter particularly decisive to deal with the QCD multijet background. We estimate the exclusion limits at 95\% C.L. and the significances for evidence and discovery at 300 fb$^{-1}$, and present these results along with those reported in a similar search by ATLAS in~\cite{Aaboud:2018jff} for an integrated luminosity of 36.1~fb$^{-1}$. As commented above, we observe a growth in the sensitivities for high Higgs boson masses at this luminosity.

Lastly, taking into account that the Higgs boson decay into a tau-lepton pair may mimic our LFV signal, we also study its impact on the exclusion limits and the significance for evidence and discovery, being this analysis not considered in the experimental searches so far. In particular, although the $H/A \rightarrow \tau\tau$ decay has not been observed, it is worth mentioning that its impact on the exclusion limits of LFV heavy Higgs boson decays becomes significant when its branching ratio begins to compete with the corresponding to the $H/A \rightarrow \tau\mu$ decay.

The paper is organized as follows: Section~\ref{sec:collider} is dedicated to the collider analysis of the $pp\to H/A \rightarrow \tau\mu$ signal, with the performance of our model-independent search strategy and the sensitivities and exclusion limits that we obtain, while Section~\ref{sec:ditau-channel} is devoted to the study of the impact of considering the Higgs ditau channel on these previous results.  Finally, we present in Section~\ref{sec:conclusions} the main conclusions of our work.

\section{Collider analysis of the $H/A\to\tau\mu$ channel}
\label{sec:collider}

In this Section we carry out the collider analysis and define our proposal of search strategies, at the next runs of the LHC, for heavy Higgs bosons decaying into a muon and a tau-lepton. In Section~\ref{sec:simulation} we explain how we have proceeded to simulate the signal and estimate the different SM backgrounds, while we present in Section~\ref{sec:SRs} the characterization of the signal that allows us to define a search strategy for three mass windows. Finally, Section~\ref{sec:LHC} is dedicated to show the results of our search strategy, with the prospects for the sensitivities that could be reached at the LHC.

\subsection{Signal simulation and background estimation}
\label{sec:simulation}

The experimental signature we are interested in consists of a final state with one tau and one muon of opposite sign  charge originated from the decay of a heavy Higgs boson.  
The signal process\footnote{We have assumed that both the CP-even and CP-odd Higgs are degenerated 
in mass. However, we have checked that similar results are obtained when 
either the CP-even or the CP-odd Higgs is decoupled. Since we have 
assumed CP conserving interactions, we have considered both channels 
$\mu {\bar \tau}$ and $\tau {\bar \mu}$ in the analysis and we denote 
them as $\tau\mu$.} $pp\to H/A \to \tau\mu$ has been simulated by means of an UFO model implemented in {\tt MadGraph\_aMC@NLO 2.6}~\cite{Alwall:2014hca}. 
In this simplified model we have included an effective coupling for the heavy Higgs boson dominant production via gluon fusion\footnote{We have found similar search strategies and sensitivities when considering the $b\bar b$ annihilation as the single production channel. This production mode is well motivated, e.g., in supersymmetry models with large $\tan\beta$ where it is the dominant production mechanism.}, as well as an effective coupling for the LFV Higgs boson decays into a tau and a muon. Both the signal and the SM backgrounds have been generated with {\tt MadGraph\_aMC@NLO 2.6}~\cite{Alwall:2014hca}, while the showering and hadronization have been performed with {\tt PYTHIA 8}~\cite{Sjostrand:2014zea}. Finally, the simulation of the detector response has been done with {\tt Delphes 3}~\cite{deFavereau:2013fsa}, where the tau-leptons are reconstructed by means of their hadronic decays. We  use the default set of parameters provided by {\tt Delphes 3} for the efficiencies and fake rates including, in particular, the miss-tagging of a jet as a tau-lepton.

The SM backgrounds for this exotic process can be sorted into a reducible category and an irreducible one. 
The irreducible category is made up of the Drell-Yan (DY) production of a tau-lepton pair, diboson production, $t \bar t$, and single-top production. 
For the diboson process, we have taken into account all the combinations of $W^\pm$ and $Z$ bosons decaying leptonically or hadronically, with at least one of the two gauge bosons decaying leptonically (muon or tau). 
The $t \bar t$ production considers both leptonic and hadronic tops, with at least one leptonic top, and the same for the single-top production, with at least the top or a $W^\pm$ boson decaying into leptons. 
On the other hand, $W+$jets and QCD multijet processes are the main reducible backgrounds, in which there are jets faking a tau and/or a muon. 
The former has been computed with matching up to two extra jets, with the $W^\pm$ boson decaying always leptonically and a jet faking an hadronic tau. 
For the QCD multijet background (matching up to two and three jets samples), we have assumed a constant fake-rate of detecting a jet as a muon of $10^{-3}$~\cite{charla:fakerate}.
We notice, however, that a more realistic treatment of the reducible backgrounds requires the use of data-driven techniques that are out of the scope of this work.

In order to optimize the simulated events in our region of interest, we have imposed the following conditions at the generator level:
\begin{align}
\label{basic-cuts}
\hspace{1.5cm}&& 
p_T^{j}&> 20 \, \text{GeV} \,, &|\eta^{j}| &< 5.0 \,,&\Delta R_{j\ell} &> 0.4 \,,
&\hspace{1.5cm} \nonumber \\
&& p_T^{\mu, \tau} &> 250 \, \text{GeV} \,, & |\eta^{\mu, \tau}| &< 2.5 \,,&\Delta R_{\mu\tau} &> 0.4 \,,
\end{align}
where $p_T^j$ and $\eta^j$ are the transverse momentum and the pseudorapidity of the jets, respectively (the same definitions apply to the muon and tau leptons), $\Delta R_{j\ell}$ is the angular distance between a jet and a lepton, and $\Delta R_{\mu\tau}$ is the angular distance between the muon and the tau.
The strong requirement over the transverse momentum of the charged leptons  is due to the fact that we are interested in the LFV decays of heavy Higgs bosons, with $M_{H,A} \gtrsim$ 1 TeV, resulting in very energetic $\tau$ and $\mu$ leptons. Therefore, these stringent cuts do barely affect the signal but reduce significantly the background cross sections and consequently the number of events to be generated.
Moreover, given the large cross section of the multijet background, we adopted the procedure developed in~\cite{Avetisyan:2013onh} to accomplish a more realistic simulation of the tail of the kinematical distributions with an achievable number of events by generating in independent bins of increasing values of the variable $H_{T2}\equiv p_T^{j_1}+p_T^{j_2}$, being $j_1$ and $j_2$ the leading and subleading jets in the event. This will allow us to obtain a reliable estimation of the acceptances corresponding to the cuts involved in our search strategies. We provide more details on the simulation of the QCD multijet background in the Appendix~\ref{appendixA}.

\renewcommand{\arraystretch}{1.2}
\begin{table}[t!]
\vspace*{4mm}
\begin{center}
\begin{tabular}{c|c|c|c}
\hline \hline
{\bf Background} & {\bf LO cross section [fb]} & {\bf K-factor} & {\bf Simulated events} \\ \hline\hline 
$W\text{+jets}$ & $4510$ & 1.6 & $2.5\times10^{6}$ \\ 
multijet & $2.3\times10^{7}$ & 1.36 & ~~~$8.3\times10^{5\,*}$ \\ 
$t\bar{t}$ & 275 & 1.5 & $1.5\times10^{5}$ \\
single-top & 77 & 1.4 & $5\times10^{4}$ \\ 
Drell-Yan & 39  & 1.4 & $2\times10^{4}$ \\ 
Diboson & 35 & 1.6 & $2\times10^{4}$ \\  \hline \hline
\end{tabular}
\caption{SM backgrounds along with the corresponding LO cross sections computed with the conditions of Eq.~(\ref{basic-cuts}) at a LHC energy of 14 TeV, K-factors extracted from~\cite{Alwall:2014hca}, and the number of simulated events.
$^*$ This number of events has been generated in exclusive bins of $H_{T2}$, as explained in the Appendix~\ref{appendixA}.
}
\label{tab_bkg}
\end{center}
\end{table}

The most important information about the generated events for the SM backgrounds is listed in Table~\ref{tab_bkg}. We estimated the cross sections at leading order (LO) by using {\tt MadGraph\_aMC@NLO 2.6} for a center-of-mass energy of $\sqrt{s} =$ 14 TeV after applying Eq.~\eqref{basic-cuts}, and then rescaled them with the corresponding K-factors extracted from~\cite{Alwall:2014hca}. We also show in Table~\ref{tab_bkg} the number of generated events for each background, which is consistent with a total integrated luminosity of ${\cal L} =$ 300 fb$^{-1}$.


\subsection{Signal characterization and search strategies}
\label{sec:SRs}

Since we are interested in the search for heavy Higgs bosons, scalar $H$ or pseudoscalar $A$ with masses $M_{H,A} \gtrsim$ 1 TeV, which decay into a muon and a hadronic-tau lepton, one expects their decay products to be very energetic.
These leptons will have a higher transverse momentum than the ones corresponding to the background processes, which mainly come from the decays of much lighter particles ($W^\pm$ and $Z$ bosons, top quarks, and leptonic taus) or misidentified light jets.
Therefore, the first step to characterize the signal, requires at the detector level one muon and one hadronic tau of opposite sign in the final state, with $p_T^\mu >$ 250 GeV and $p_T^{\tau\text{-vis}} >$ 250 GeV. 
This requirement is consistent with the conditions at the generator level in Eq.~\eqref{basic-cuts}.

In summary, collecting all the requirements we mentioned so far,
\begin{align}\label{cuts-charac}
&N_{\mu}=N_\tau =1\,, && Q_\mu\cdot Q_\tau < 0 \,, && |\eta^{\mu,\tau}|<2\,,
&p_T^{\mu, \tau\text{-vis}} > 250~{\rm GeV}\,,
\end{align}
where $N_{\mu (\tau)}$ is the number of muons (taus) and $Q_{\mu (\tau)}$ is the electric charge of the muon (tau). We have considered as tau candidates those with only one charged track, since the jets misidentified as hadronic taus tend to have associated more than one charged track.
Moreover, we veto any event with electrons or bottoms, and require a stronger $\eta$ selection than in Eq.~\eqref{basic-cuts}, which barely affects the signal while avoiding values of $\eta$ for which the jet missidentification rate could be higher~\cite{charla:fakerate}.
Notice also that, since our search strategy is performed by taking the detector effects into account, the tau-leptons are only partially reconstructed via their visible (hadronic) component, with the associated neutrino contributing to the $E_T^\text{miss}$.
Nevertheless, in order to simplify the notation, we will use from now on the notation $p_T^\tau$ to refer to the reconstructed visible momentum $p_T^{\tau\text{-vis}}$.

\begin{figure}[t!]
\begin{center}
\includegraphics[width=.49\textwidth]{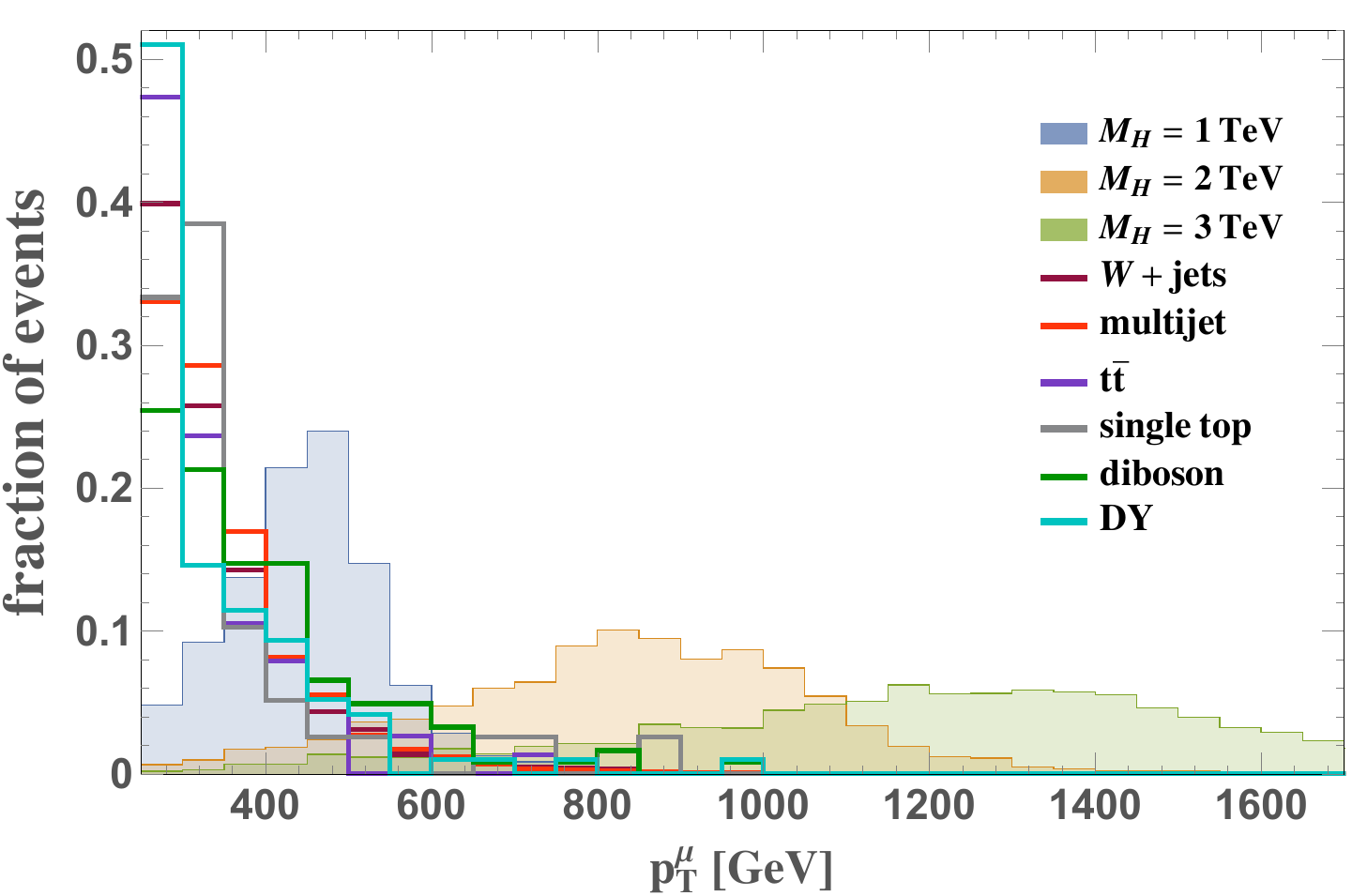} 
\includegraphics[width=.49\textwidth]{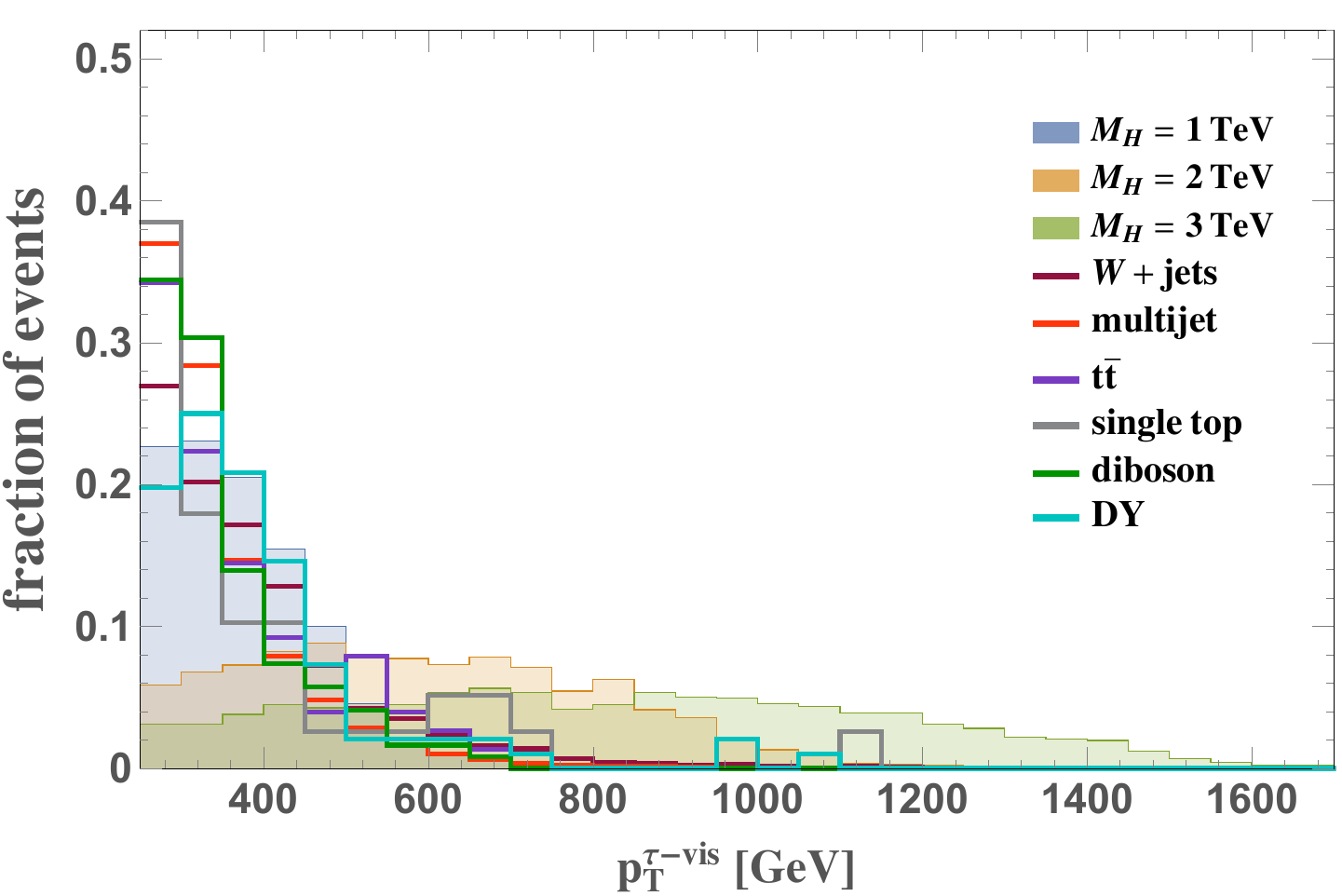}
\caption{Distribution of the transverse momentum of the muon (left panel) and the visible hadronic-tau lepton (right panel), for three signal benchmarks ($M_H =$ 1 TeV, 2 TeV, and 3 TeV) and the main SM backgrounds, after minimal requirements in Eq.~\eqref{cuts-charac}, for a LHC center-of-mass energy of $\sqrt{s} =$ 14 TeV.}\label{pT-distributions}
\end{center}
\end{figure}

After having characterized the signal, we can search for the best cuts to favor further the signal over the background. 
The main idea is to exploit the fact that the Higgs boson is heavy, from which it is expectable to have decay products with high momenta, peaked approximately at $p_T^\mu \sim$ $p_T^\tau \sim$ $M_H/2$. 
This is seen in Fig.~\ref{pT-distributions}, where we display the profile of the distributions of the transverse momentum of the muon (left panel) and the visible hadronic-tau lepton (right panel), for three signal benchmarks ($M_H =$ 1 TeV, 2 TeV, and 3 TeV) and the  SM backgrounds, with a center-of-mass energy of $\sqrt{s} =$ 14 TeV. 
The $p_T^\mu$ distributions of all the background\footnote{In the case of the multijet background, we consider the leading jet as a fake muon for the distributions. Additional configurations with subleading jets faking the muon will be removed by the requirement $p_T^\mu > p_T^j$ in Eq.~\eqref{search-strategy_cuts}.} processes are concentrated on values below 400 GeV, peaked in most of the cases on the first bins of $p_T^\mu$, whilst the distributions for the three signal benchmarks depicted here are peaked for high values of $p_T^\mu$, following the Higgs boson mass hypothesis. 
It is patent then that a strong cut on the transverse muon momenta should be very helpful to increase the signal-to-background ratio, as we will discuss later. 
On the other hand, we cannot obtain the same conclusion from the $p_T^\tau$ distributions, in which the signal $p_T^\tau$ profiles do not reach values as large as the $p_T^\mu$ ones. The reason is that the $p_T^\tau$ is reconstructed from the visible hadronic tau but an important fraction of the total transverse tau-lepton momentum is carried by the invisible tau-decay product, the corresponding neutrino.
Nevertheless, the latter will lead to an important amount of missing transverse energy, $E_T^\text{miss}$, which can also be used to characterize our signal.

\begin{figure}[t!]
\begin{center}
\includegraphics[width=.49\textwidth]{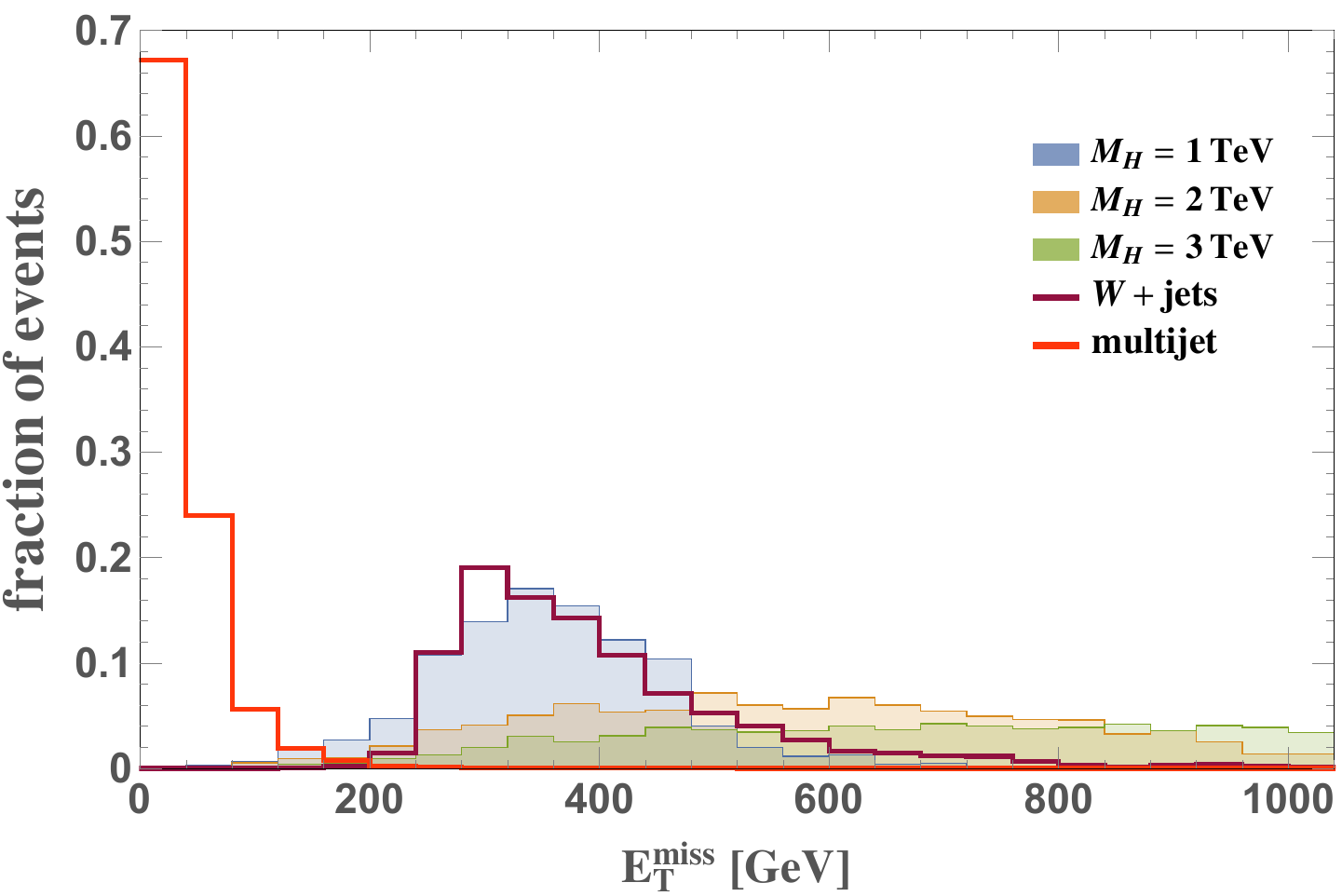}
\includegraphics[width=.49\textwidth]{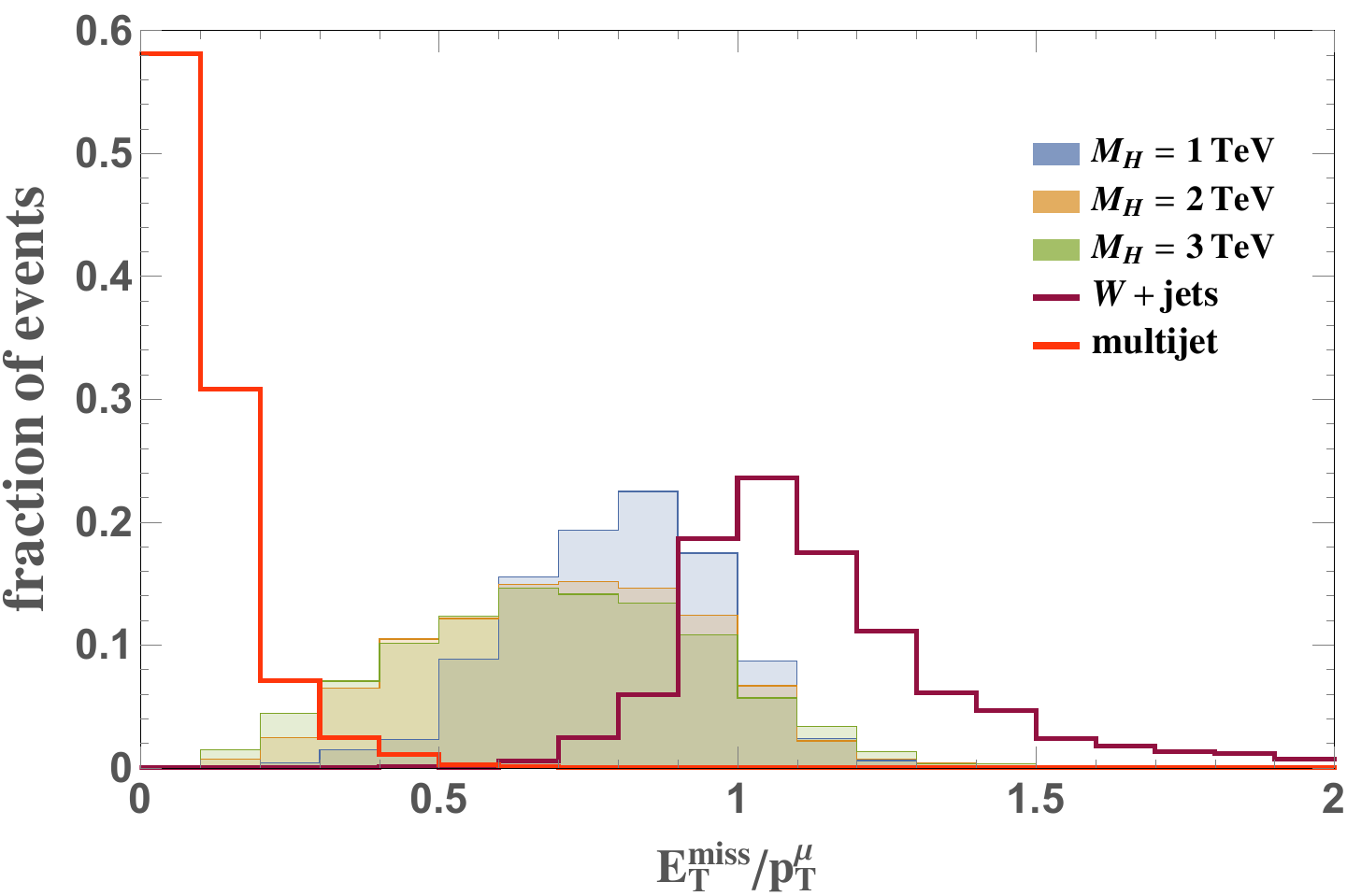}
\caption{Distribution of the missing transverse energy $E_T^\text{miss}$ (left) and $E_T^\text{miss}/p_T^\mu$ (right) for three signal benchmarks ($M_H =$ 1 TeV, 2 TeV, and 3 TeV) and the main reducible backgrounds $W+$jets and multijet, after minimal requirements in Eq.~\eqref{cuts-charac}, for a LHC center-of-mass energy of $\sqrt{s}=$~14 TeV. }\label{MET-distributions}
\end{center}
\end{figure}

We display the  missing transverse energy in Fig.~\ref{MET-distributions} for the three signal benchmarks as well as for the two reducible backgrounds. 
For simplicity, we show only these backgrounds as they are the dominant ones after minimum requirements of Eq.~\eqref{cuts-charac}.
We clearly see that the multijet background is peaked close to zero, as expected since it does not possess a genuine source of $E_T^\text{miss}$.
Therefore, a lower cut on the $E_T^\text{miss}$ will be useful to suppress this background. 
On the other hand, in both the signal and the $W+$jets background there is a real source of $E_T^\text{miss}$ from the neutrinos of the hadronic-tau leptons and the leptonic decays of $W^\pm$ bosons.
Nevertheless, these two cases can be partially decorrelated by studying the ratio $E_T^\text{miss}/p_T^{\mu}$, as shown in the right of Fig.~\ref{MET-distributions}.
The three signals studied here tend to lead to values smaller than one, since the only source of missing transverse momentum comes from the tau-lepton decay with $p_T^\tau\sim p_T^\mu$. 
This is not necessarily the case for  the $W+$jets background, therefore requiring that $E_T^\text{miss} <p_T^\mu$ will help reducing it. 
Indeed, and more generally,  what we will want to do is to look for events in which the leading particle is a muon, in a similar way to the dynamical jet veto proposed in~\cite{Pascoli:2018rsg}.

In summary, we want to design a search strategy that selects events where the muon is the leading particle and imposes lower cuts for $p_T^\mu$, $p_T^\tau$ and $E_T^\text{miss}$. 
As we see in Figs.~\ref{pT-distributions} and \ref{MET-distributions}, the heavier the Higgs boson, the stronger the cuts we could impose. 
In this sense, designing a strategy with very high-$p_T$ and $E_T^\text{miss}$ cuts would improve the sensitivities in the heavy $M_H$ regime, however it would kill the sensitivities for lower masses. 
On the contrary,  lowering the cuts to explore a wider range of $M_H$ would not be as efficient as possible at high masses. 
And interesting compromise between the two is to design a strategy that varies the cuts according to the tested heavy mass hypothesis.
Therefore, keeping that in mind, we found that the following set of kinematic cuts, on top of Eq.~\eqref{cuts-charac}, improves the sensitivity for each hypothesis of the heavy Higgs boson mass:
\vspace*{2mm}
\begin{align}
\label{search-strategy_cuts}
& p_T^\mu/M_H > 0.4 \,, 
&& p_T^\tau/M_H > 0.1\,,
&& E_T^\text{miss}/M_H > 0.25\,,
&& p_T^\mu \geq p_T^\tau, p_T^{j}, E_T^{\rm miss}\,.
\end{align}
In principle, one could apply the criteria in Eq.~\eqref{search-strategy_cuts} for each hypothesis of $M_H$, defining a {\it dynamical} set of cuts. 
Nevertheless, we split our search strategy in three $M_H$ mass windows: [1, 1.5) TeV, [1.5, 2.5) TeV and [2.5, 5] TeV, denoted as {\it low}, {\it medium} and {\it high} from now on.
In each of these mass windows, we use Eq.~\eqref{search-strategy_cuts} with reference values\footnote{We have also explored setting the reference value for the {\it high} region to a higher value. Nevertheless we found similar results, since the cuts for $M_H=2.5$~TeV already remove almost all the background events for $\mathcal L=300~{\rm fb}^{-1}$. This might not be true for higher luminosities. } for $M_H$ of 1, 1.5 and 2.5~TeV, respectively. 
We summarize the three analysis explicitly in Table~\ref{tab_SRs}, as well as the chosen $M_H$ region for each case \footnote{Observe that the lower cut on $p^{\tau}_T$ is the same for the three mass windows in spite of the condition imposed by the Eq.~\eqref{search-strategy_cuts} since the events have been generated with a minimum value $p^{\tau}_T = 250$~GeV (see Eq.~\eqref{basic-cuts}).}.

\begin{figure}[t!]
\begin{center}
\includegraphics[width=.48\textwidth]{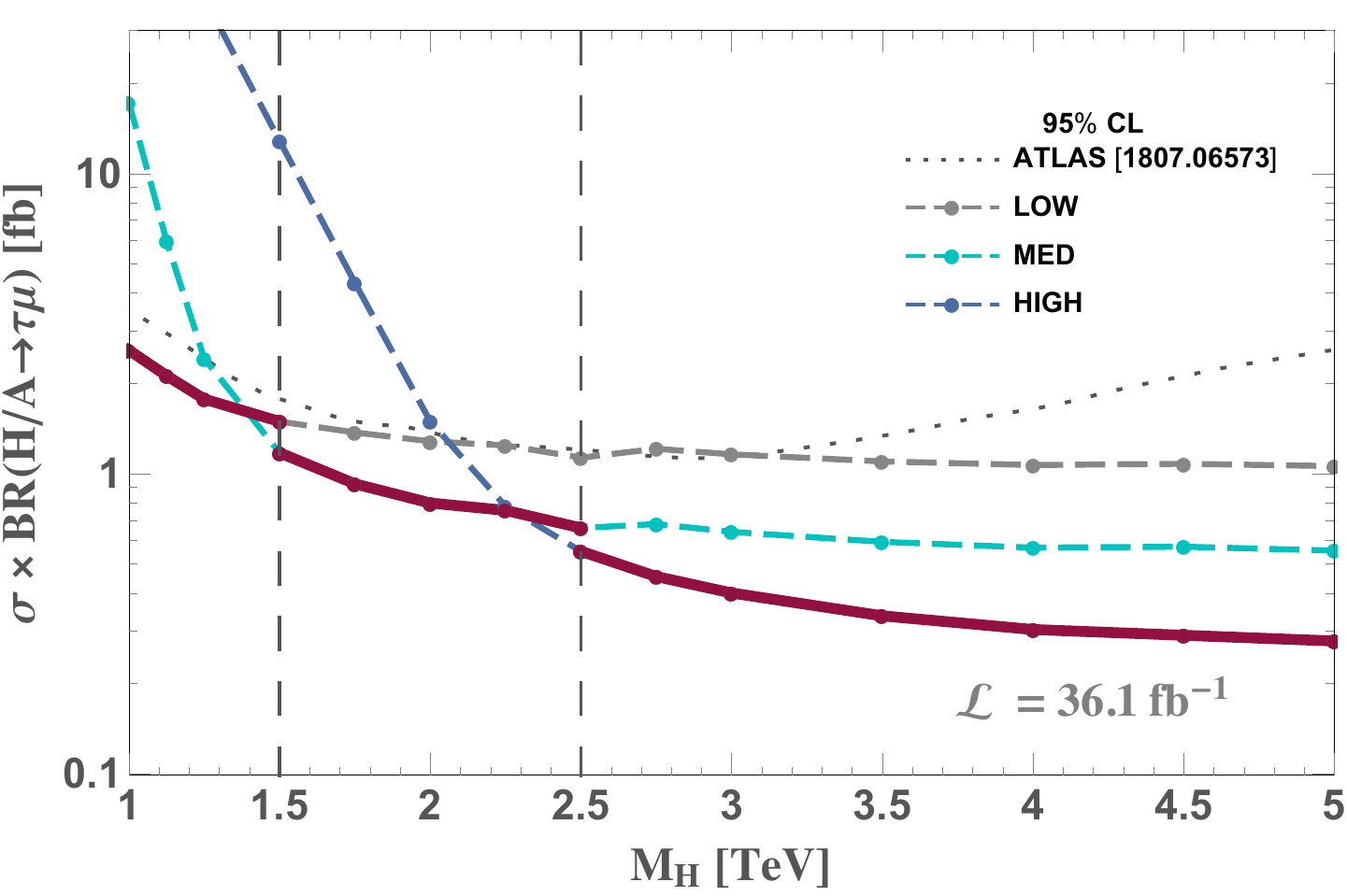} 
\caption{95\% C.L. exclusion limits in the [$M_H$, $\sigma(pp \to H/A \to \tau\mu)$] plane  for the search strategies {\it low} (gray dashed line), {\it medium} (cyan dashed line), {\it high} (blue dashed line), and our global search strategy in the full mass range (maroon solid line), with ${\cal L} =$ 36.1 fb$^{-1}$. 
Dashed vertical lines indicate the mass region for each analysis, see Table~\ref{tab_SRs}.
The dotted line is the expected exclusion for a sneutrino $\widetilde\nu \to \tau\mu$ search by ATLAS at ${\cal L} =$ 36.1 fb$^{-1}$~\cite{Aaboud:2018jff}.}\label{95CL_36invfb}
\end{center}
\end{figure}

\renewcommand{\arraystretch}{1.2}
\begin{table}[t!]
\begin{center}
\begin{tabular}{c|c|c|c}
\hline \hline
{\bf Analysis} & {\bf LOW} & {\bf MEDIUM} & {\bf HIGH} \\ \hline\hline 
$M_H$ region [TeV] & [1, 1.5) & [1.5, 2.5) & [2.5, 5] \\ 
$p_T^\mu$ [GeV] & $> 400$ & $> 600$ & $> 1000$ \\
$p_T^\tau$ [GeV] & $250<p_T^\tau<p_T^\mu$ & $250<p_T^\tau<p_T^\mu$ & $250<p_T^\tau<p_T^\mu$ \\
$E_T^\text{miss}$ [GeV] & 250 $< E_T^\text{miss} <$ $p_T^\mu$  & 375 $< E_T^\text{miss} <$ $p_T^\mu$ & 625 $< E_T^\text{miss} <$ $p_T^\mu$  \\  \hline \hline
\end{tabular}
\caption{Definition of our analysis, which is split in three different search strategies depending on the $M_H$ hypothesis. The cuts in each regime follow Eqs.~\eqref{cuts-charac} and \eqref{search-strategy_cuts}.}
\label{tab_SRs}
\end{center}
\end{table}
%
We show in Fig.~\ref{95CL_36invfb} the 95\%~C.L. exclusion limits, computed as explained below, following the defined {\it low}, {\it medium} and {\it high} strategies for $\mathcal L = 36.1~{\rm fb}^{-1}$. 
The reason to choose this integrated luminosity is also to show the sensitivity that ATLAS expected in a similar search for LFV sneutrino decays $\widetilde \nu\to \tau\mu$~\cite{Aaboud:2018jff} as a qualitative reference for our strategy.
Our three analysis show the above discussed behavior: the {\it low} strategy gives the best results for lower masses, however it is less efficient for higher masses. 
This is, to some extent, what also happens to the analysis performed by ATLAS.
On the other hand, the {\it high} analysis is very efficient at high masses, improving the reach of the {\it low} analysis in one order of magnitude, nevertheless it is not competitive at lower $M_H$. 
For the same reasons, the {\it medium} analysis covers more efficiently intermediate mass hypothesis. 
The crossings between the three analysis happen approximately at masses of 1.5 and 2.5~TeV, motivating thus our choice of $M_H$ regions in the definition of our search strategy in Table~\ref{tab_SRs}.
This is depicted with a maroon solid line in Fig.~\ref{95CL_36invfb}, showing that it combines the best qualities of each analysis. 
Consequently, from now on we will use our search strategy by mass windows to explore the future sensitivities. 

As we can see from Fig.~\ref{95CL_36invfb}, for Higgs boson masses above 2.5 TeV the ATLAS analysis loses sensitivity due to a decrease in acceptance at very high $p_T$~\cite{Aaboud:2018jff}. However, with improvements in the tau reconstruction at high $p_T$~\cite{Bha:2019}, it may be expected that the experimental limits follow a tendency similar to the one we obtain in the {\it high} mass window under the assumption that the tau reconstruction is not deteriorated for the high $p_T$ values associated to this window.

\subsection{LHC sensitivities and exclusion limits}
\label{sec:LHC}

After introducing our search analysis in the previous Subsection, we can now compute and discuss the potential of the future LHC runs for exploring the LFV decays of heavy Higgs bosons. 

Our statistical analysis is based on the tests statistic $q_0$ and $q_\mu$~\cite{Cowan:2010js}. In order to establish 95\%~C.L. exclusion limits, we consider the p-value of the $q_\mu$ test (corresponding to the only-background hypothesis) lower than 0.05 and in terms of the significance:
\begin{equation}
   {\cal S}_{\rm excl}= \sqrt{2 \left(B \log \left(\frac{B}{B+S}\right)+S\right)} \leq 1.64\,,
    \label{exclusion95CL-formula}
\end{equation}
where $S$ and $B$ are the number of signal and background events at a given luminosity $\cal{L}$. Using $S=\sigma(pp \to H/A \to \tau\mu) \times \cal{L}$, we obtain the exclusion limits on $\sigma(pp \to H/A \to \tau\mu)$ for $\cal{L}=$ 36.1~fb$^{-1}$ showed in Fig.~\ref{95CL_36invfb}.

The evidence/discovery sensitivities are obtained from the $q_0$ test given by
\begin{equation}
    {\cal S} = \sqrt{2 \left((B+S) \log \left(\frac{S+B}{B}\right) -S \right)} \,.
    \label{Snosys}
\end{equation}
The evidence (3$\sigma$) and discovery (5$\sigma$) sensitivities correspond to the constraints ${\cal S} \leq$ 3 and 5, respectively, and we set limits on $\sigma(pp \to H/A \to \tau\mu)$ for a given luminosity $\cal{L}$.

\begin{figure}[t!]
\begin{center}
\includegraphics[width=.48\textwidth]{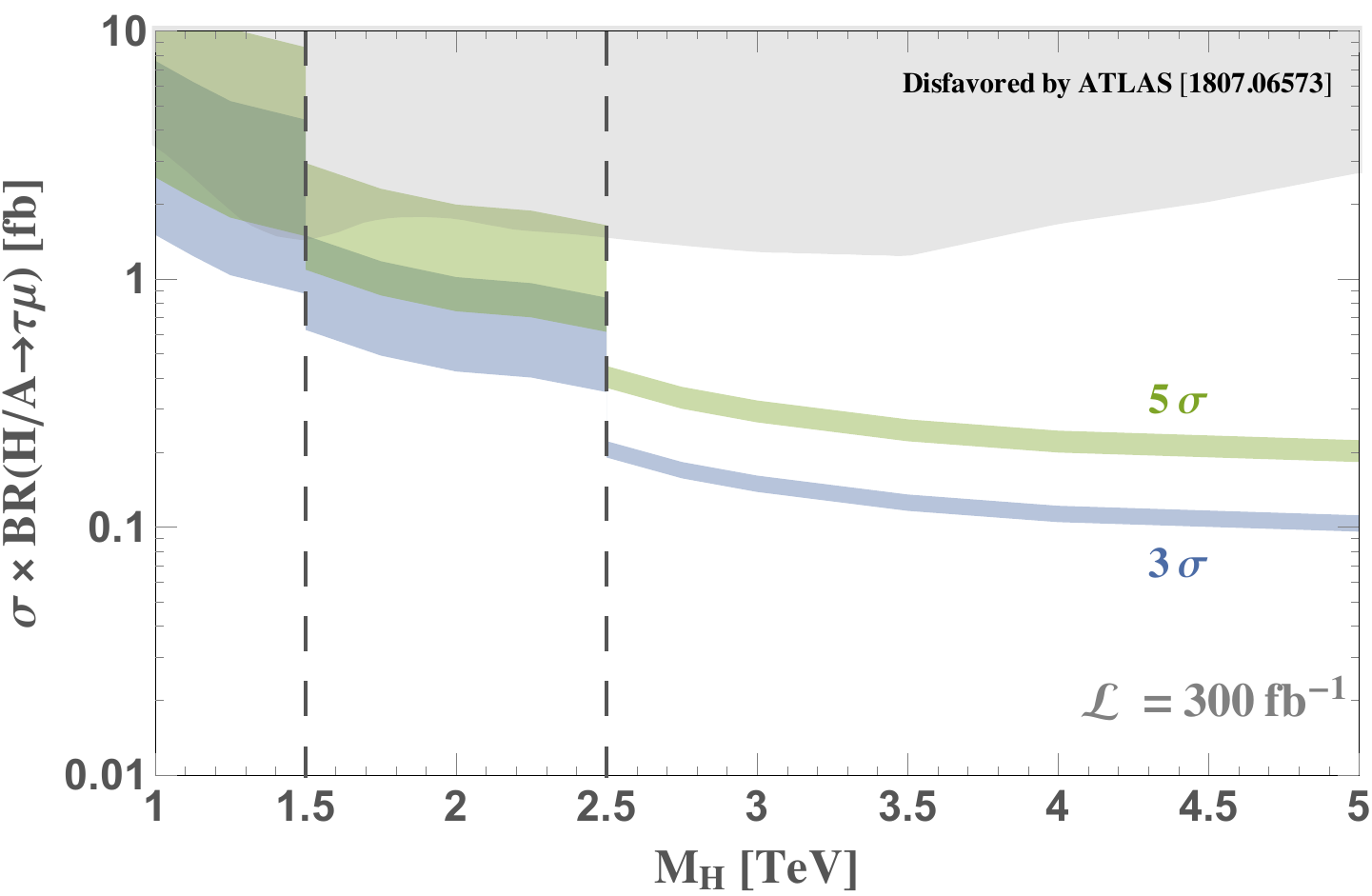}
\includegraphics[width=.48\textwidth]{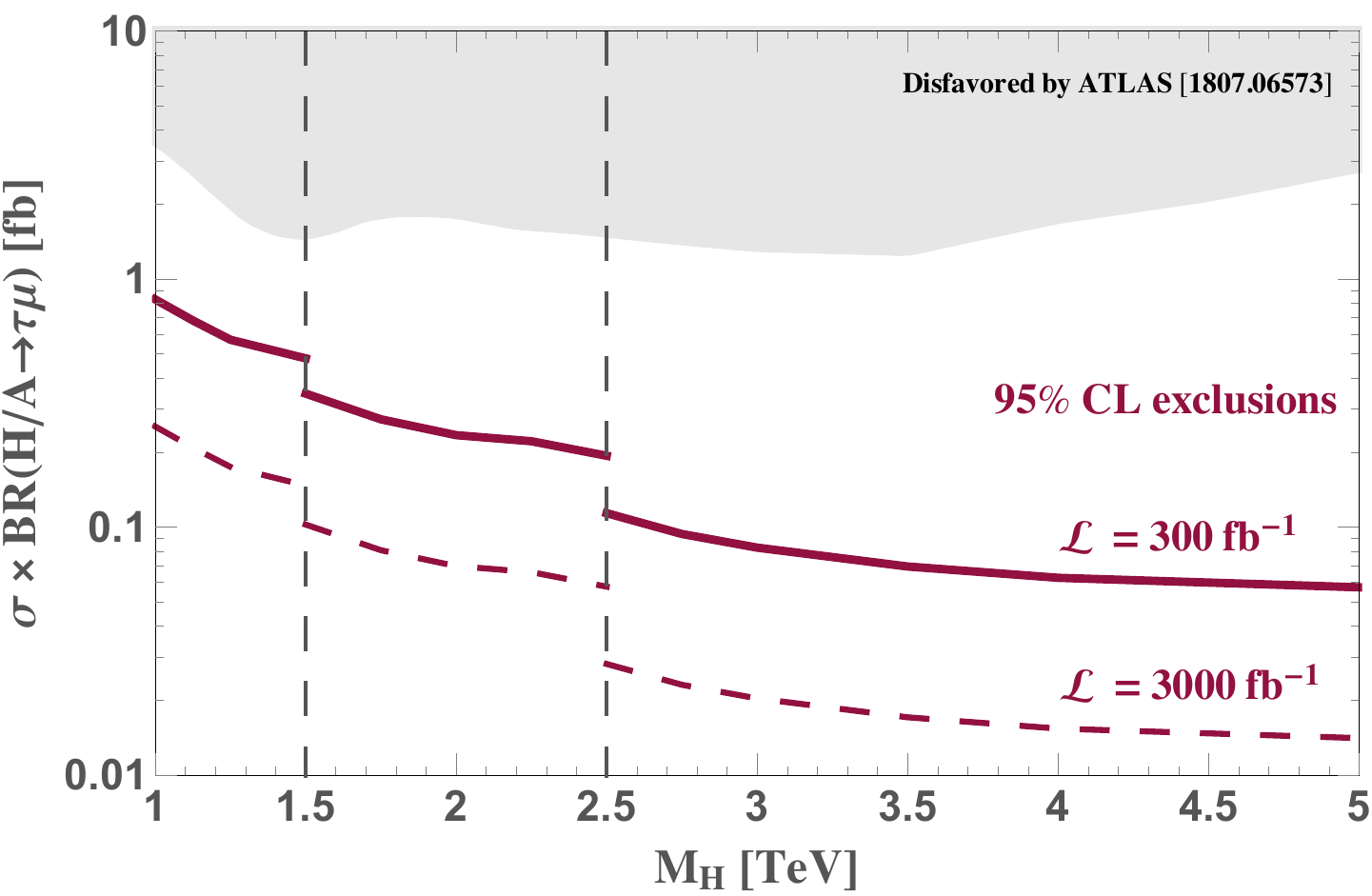} 
\caption{Left: 3$\sigma$ (blue) and 5$\sigma$ (green) significances with ${\cal L} =$ 300 fb$^{-1}$. 
The shadowed contours over the lines show the effect of adding up to a 30\% of systematic errors. Right: 95\% exclusion limits with ${\cal L} =$ 300 fb$^{-1}$ (solid red line) and scaled, as explained in the text, to ${\cal L} =$ 3000 fb$^{-1}$ (dashed red line).  
Dashed vertical lines indicate the mass region for each analysis, see Table~\ref{tab_SRs}.
The gray area is disfavored at 95\% CL by the ATLAS similar search for LFV high-mass final states in the $\widetilde \nu\to\tau\mu$ channel at ${\cal L} =$ 36.1 fb$^{-1}$~\cite{Aaboud:2018jff}.} \label{95CL3-5sigma_300invfb}
\end{center}
\end{figure}

We display in the right panel of Fig.~\ref{95CL3-5sigma_300invfb} the 95\% C.L. exclusion limits for ${\cal L} =$ 300 fb$^{-1}$ (solid line) corresponding to our search strategy of Table~\ref{tab_SRs}. We separate the three mass regions ({\it low}, {\it medium} and {\it high}) using the dashed vertical lines. For the lighter Higgs boson masses, we can excluded cross sections of $\cal{O}$(1) fb but for heavier Higgs boson masses this search strategy is sensible to $\cal{O}$(0.1)~fb. We assumed that the number of background events scales as the signal one, so we extrapolated our exclusion limits to ${\cal L} =$ 3000 fb$^{-1}$ (dashed line).

On the other hand, we present the evidence (blue band) and discovery (green band) sensitivities for ${\cal L} =$ 300 fb$^{-1}$ with a relative systematic uncertainty up to 30\%~\cite{Cowan:2010js} in order to give more realistic estimations with our search strategy. As heavier the Higgs boson mass hypothesis, less background events survives our search strategy, then the uncertainty band is thinner for the {\it high} mass window compared to the others. In particular, the {\it low} mass hypothesis is very constrained by the current bounds, specially taking into account the systematic uncertainties. This situation is slightly better for the {\it medium} hypothesis and, as we explained before, the best sensitivity is obtained in the {\it high} mass window.

\section{Impact of the heavy Higgs ditau channel}
\label{sec:ditau-channel}

\begin{figure}[h!]
\begin{center}
\includegraphics[width=.46\textwidth]{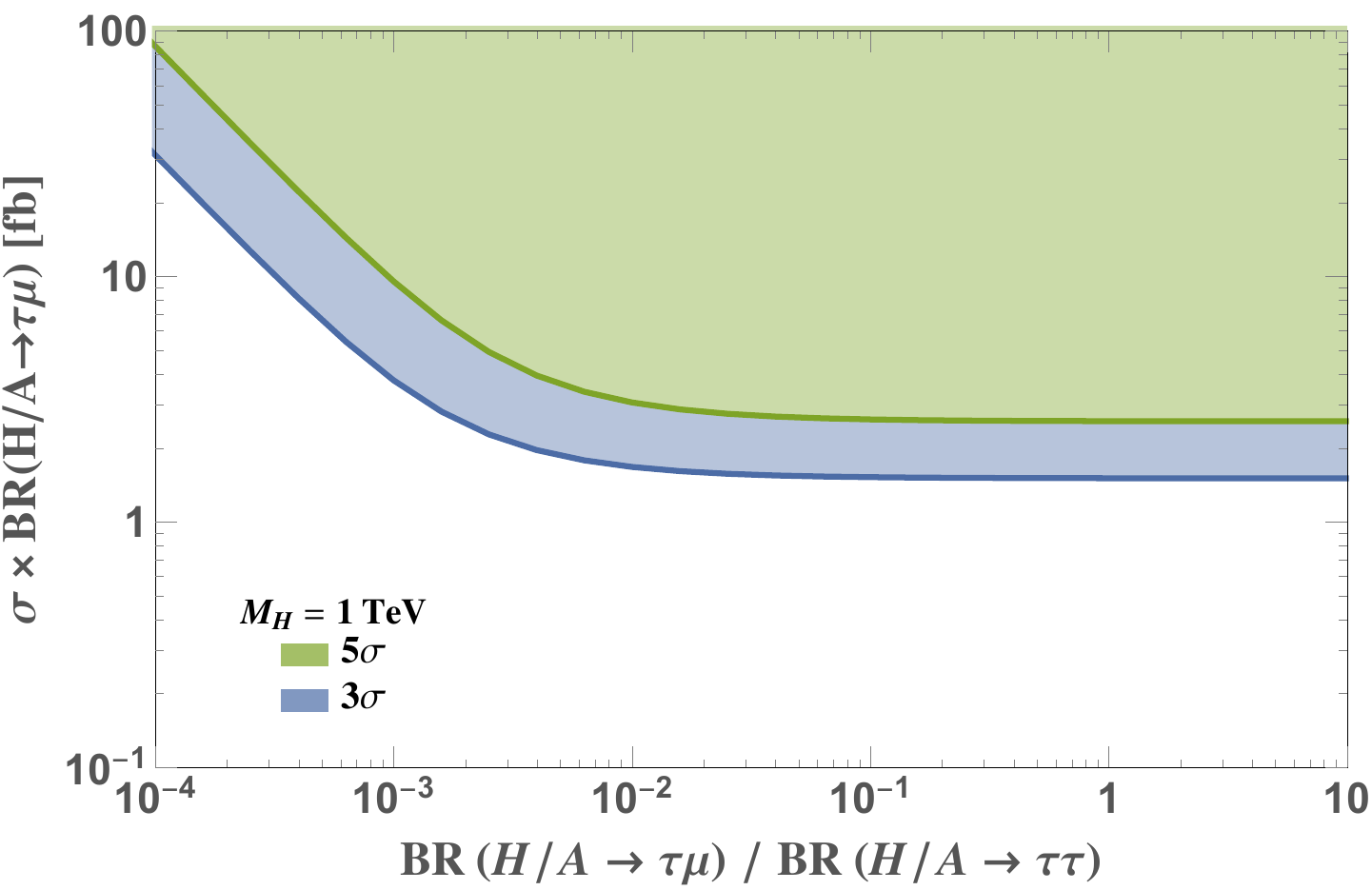} 
\includegraphics[width=.46\textwidth]{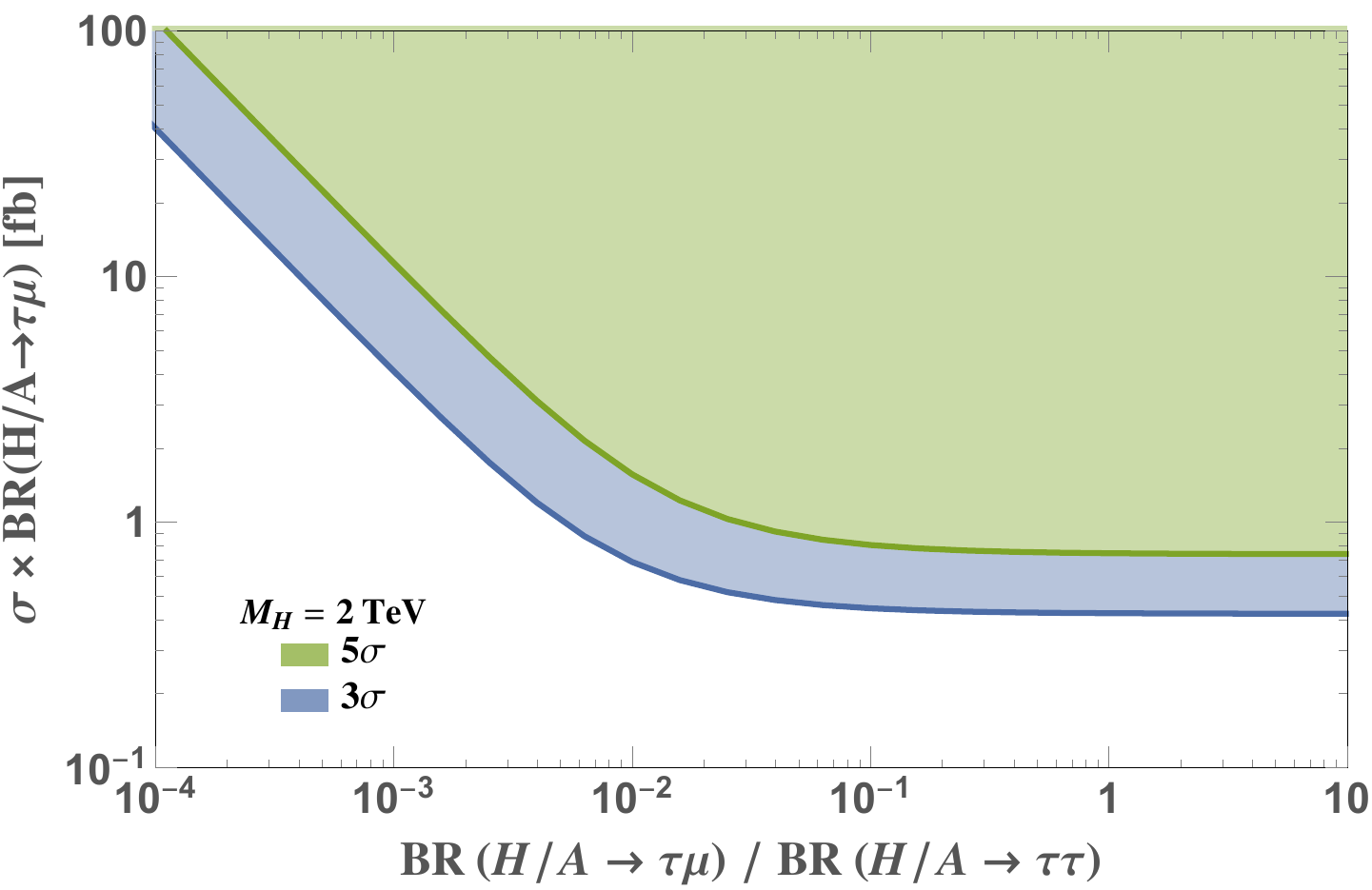}\\[4mm]
\includegraphics[width=.46\textwidth]{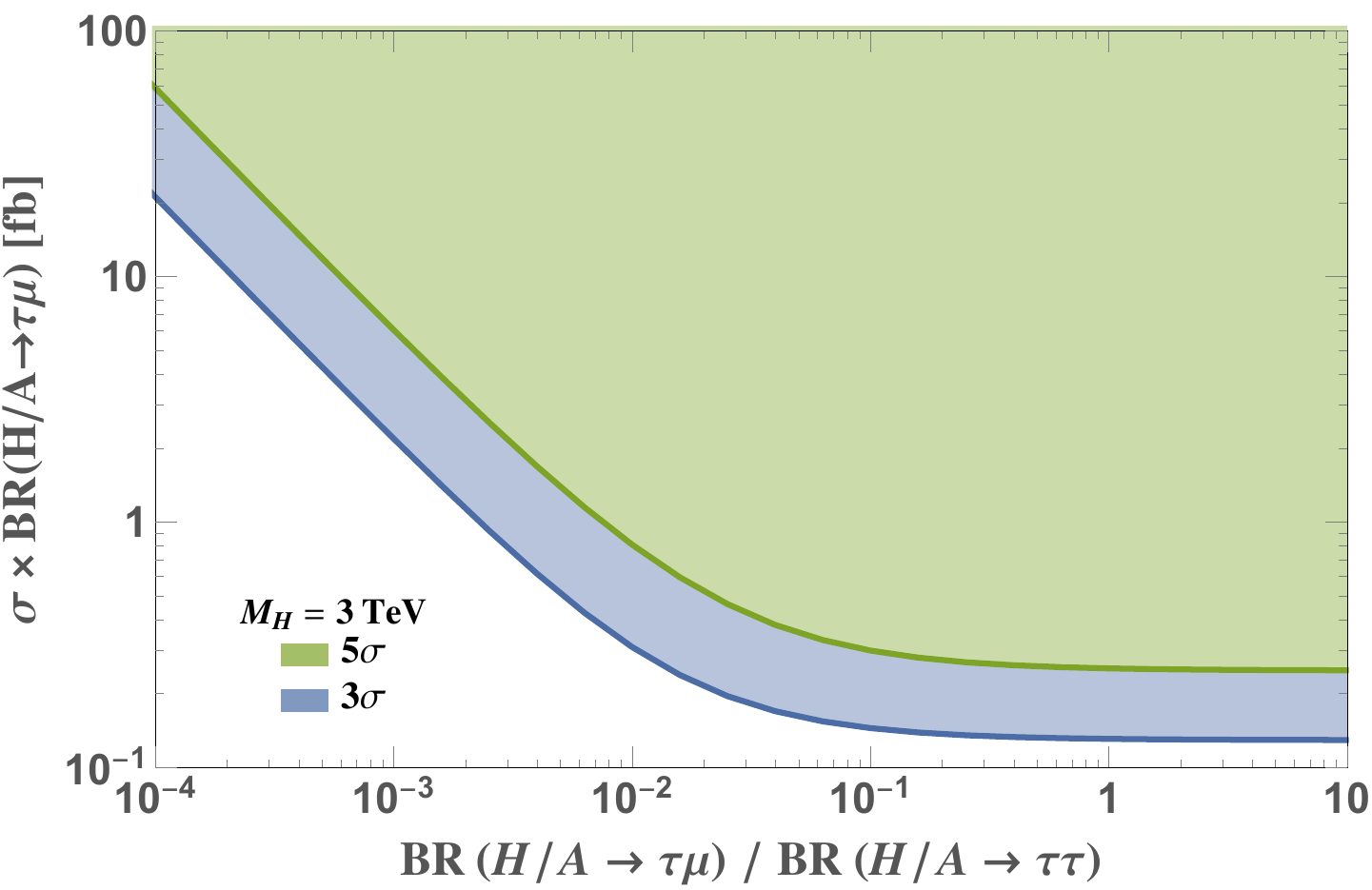}
\caption{3$\sigma$ (blue) and 5$\sigma$ (green) significances obtained with ${\cal L} =$ 300 fb$^{-1}$ for $M_H=1$ TeV (upper left), 2 TeV (upper right) and 3 TeV (bottom) when the ditau channel is included as  background.}
\label{95CL3-5sigma_ditau}
\end{center}
\end{figure}

So far we have assumed that only the LFV process $pp\to H/A \to \tau\mu$ contributes to the signal events in the three signal regions defined by the {\it low}, {\it medium} and {\it high} search strategies described previously. However, if the heavy Higgs bosons can decay into $\tau\tau$, the process $pp\to H/A \to \tau\tau$ also contributes since either of the tau-leptons could decay leptonically giving rise to a muon lepton in the final state. Given that the production cross section of $H/A$ is common to both processes, the relative strength between the cross sections of the LFV and the lepton flavor conserving (LFC) processes will be determined by the ratio $R\equiv \mathrm{BR}(H/A\to \tau\mu)/\mathrm{BR}(H/A\to\tau\tau)$. Now, a potential observation of the heavy Higgs bosons in the ditau channel would certainly impact on the capability of the search strategies developed in Section~\ref{sec:SRs} to detect the LFV process. In order to quantify this, we have computed the evidence and discovery sensitivities to the LFV process incorporating the ditau process as an additional background. The results are depicted in Fig.~\ref{95CL3-5sigma_ditau}, where we plot the $3\sigma$ and $5\sigma$ significances in terms of the ratio $R$ for three benchmark masses of the {\it low}, {\it medium} and {\it high}  search strategies. After imposing the cuts of Eqs.~(\ref{cuts-charac}) and~(\ref{search-strategy_cuts}), the acceptances of the ditau channel with respect to the LFV channel are smaller by a factor of 0.009 ({\it low}), 0.02 ({\it medium}), and 0.009 ({\it high}).

As expected, the sensitivity worsens with decreasing values of $R$. More precisely, the impact of the ditau contribution becomes significant at different values of $R$ depending on the heavy Higgs boson mass. For $M_H=1$ TeV ({\it low} search strategy), the branching ratio of the ditau channel needs to be at least two orders of magnitude higher than the branching ratio of the LFV channel, while for $M_H=2$ and $3$ TeV ({\it medium} and {\it high} search strategies, respectively) the sensitivity already worsens when the ditau branching ratio is $50$ and $10$ times the LFV branching ratio, respectively. This behavior was expected, as the number of events corresponding to the SM backgrounds is reduced by approximately two orders of magnitude from the {\it low} to the {\it high} search strategy, making the latter more sensitive to the ditau contribution. \par

\begin{figure}[t!]
\begin{center}
\includegraphics[width=.49\textwidth]{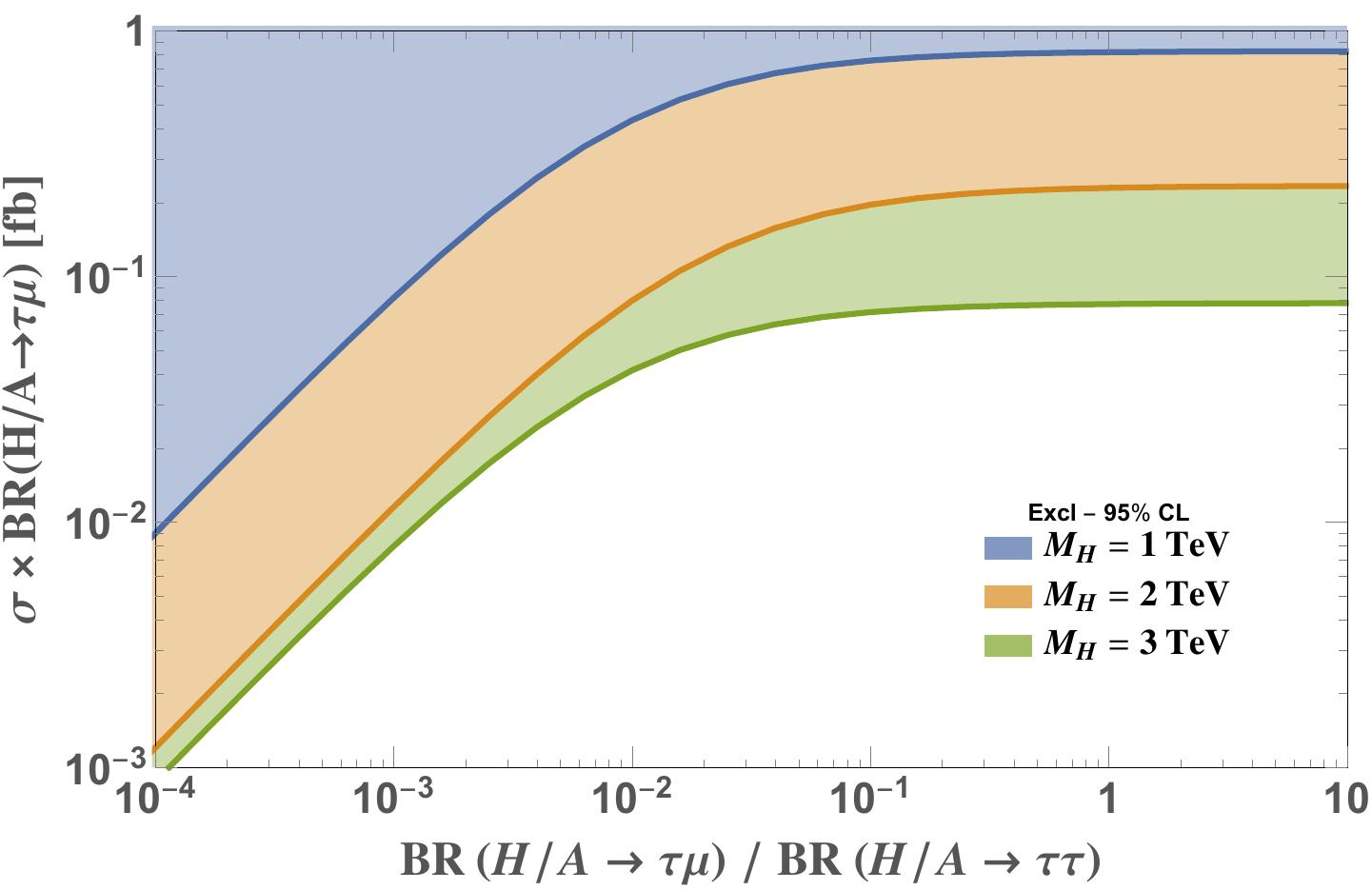} 
\caption{95\% exclusion limits on the cross section of the LFV process $pp\to H/A\to \tau\mu$ when the LFC decay $H/A\to \tau\tau$ is available. The three contours are obtained for ${\cal L} =$ 300 fb$^{-1}$ and correspond to $M_H=1$ TeV (blue), 2 TeV (orange) and 3 TeV (green).}
\label{95CL_ditau}
\end{center}
\end{figure}

Let us now consider the impact of the ditau contribution on the exclusion limits imposed  on the cross section of the LFV process $pp\to H/A\to \tau\mu$. In general, in order to set exclusion bounds on a certain model all the contributing new-physics processes should be included in the signal, while the total background will arise exclusively from all the relevant SM processes. Therefore, when the process $pp\to H/A\to \tau\tau$ is also possible, its contribution must be added to the signal in order to obtain the exclusion limits on the cross section of the LFV channel. 
The results are presented in Fig.~\ref{95CL_ditau}. 
Starting from $R\sim 10^{-2}$,  a decrease of one order of magnitude in $R$ translates into an improvement of the exclusion limit of one order of magnitude. This is due to the fact that the number of signal events associated to the ditau channel that survives the cuts becomes even more important as $\mathrm{BR}(H/A\to \tau\tau)$ increases with respect to $\mathrm{BR}(H/A\to \tau\mu)$. In particular, when $R$ is lower than $\sim 10^{-1}$, $\mathrm{BR}(H/A\to \tau\tau)$ turns out to be at least one order of magnitude above $\mathrm{BR}(H/A\to \tau\mu)$, which tends to compensate the smaller acceptance of the LFC channel, thus pushing the exclusion limits to smaller values of $\sigma\times \mathrm{BR}(H/A\to \tau\mu)$. 
More precisely, for both $M_H=1$ and 3 TeV the improvement becomes important at $R\sim10^{-2}$, while $R\sim 10^{-1}$ is enough in the case of $M_H=2$ TeV due to the fact that the {\it medium} search strategy is the one with the highest acceptance for the ditau channel. 
From this discussion, we conclude that the derived bounds on the LFV process could be more stringent for models in which the heavy Higgs bosons can decay into both $\tau\mu$ and $\tau\tau$ channels than for models where only the LFV decay is present. 

\section{Conclusions}
\label{sec:conclusions}
In this work we have developed a search strategy at the LHC for heavy Higgs bosons decaying into a tau and a muon leptons, which shows a plausible improvement in the sensitivities of the current experimental bounds.
We have worked in a model-independent way with generic lepton-flavor-violating effective interactions for the heavy Higgs bosons. Our signal process corresponds to $pp\to H/A \to \tau\mu$ produced via gluon fusion as the dominant channel but we have also found similar results considering instead the $b\bar{b}$ annihilation as the dominant production mechanism. 

We optimized the search in three Higgs boson mass windows within the range 1-5 TeV, exploiting the high transverse momenta of the final charged leptons and the missing transverse energy in order to maximize the significances at $\sqrt{s}$ = 14 TeV with $\cal{L}$ = 300 fb$^{-1}$.
We have found promising improvements in the present experimental sensitivities for masses above 2.5~TeV, even when the systematic uncertainties are included in the analysis.

Finally, we have discussed the role of the $H/A\to\tau\tau$ decay channel in the LFV searches, exploring its impact on the discovery potential and the exclusion limits. 
We have found that sizable effects start to appear when the BR($H/A\to\tau\tau$) is approximately ten times larger than the BR($H/A\to \tau\mu$). In particular, the excluded regions for the LFV process become more stringent when both decay channels are present.

\section*{Acknowledgments}
X.M. would like to thank Claudia Garc\'ia-Garc\'ia for fruitful discussions.
This work has been partially supported by CONICET and ANPCyT under projects PICT 2016-0164 (E. A., N. M., A. S.), PICT 2017-2751 (E. A., N. M., A. S.) and PICT 2017-2765 (E. A.). This work is supported by the European Union through the ITN ELUSIVES H2020-MSCA-ITN-2015//674896 and the RISE INVISIBLE-SPLUS H2020-MSCA-RISE-2015//690575, by the CICYT through the project FPA2016-78645-P, and by the Spanish MINECO’s ``Centro de Excelencia Severo Ochoa'' Programme under grant SEV-2016-0597. 
\appendix
\section{Simulation of the QCD multijet background}
\label{appendixA}
In this appendix we outline the procedure we have used for simulating the QCD multijet background. Since the cross section corresponding to this background process is huge, the generation of a sample of events that accurately model the kinematic distributions for a luminosity of $300$~fb$^{-1}$ is not achievable through the use of unweighted events and it is necessary to rely on an approach based on event weighting. Following the technique developed in~\cite{Avetisyan:2013onh}, we have first computed the differential cross section with respect to the variable $H_{T2}=p_T^{j_1}+p_T^{j_2}$, being $j_1$ and $j_2$ the leading and subleading jets in the event. In order to do that, we obtain the cross section with {\tt MadGraph\_aMC@NLO 2.6} imposing a lower cut in $H_{T2}$ which is progressively increased in steps of 100 GeV, starting from 500 GeV. Remember that two jets must be reconstructed as a muon and a tau under the $p_T$ requirements of Eq.~(\ref{basic-cuts}).\par
Once the estimation of the differential cross section is complete, the bins of $H_{T2}$ for event generation are chosen in such a way that approximately one decade of cross section falls in each bin. The last bin is inclusive and determined by requiring that $\mathcal{L}\times\sigma(\mathrm{bin}_f)<N/5$, where $\sigma(\mathrm{bin}_f)$ is the cross section corresponding to the last bin ($\mathrm{bin}_f$) and $N$ is the number of events to be generated in this final bin. With this procedure we found that six bins of $H_{T2}$ are required. In each of these six bins we generate $10^6$ parton-level events at leading order, which are subsequently showered and matched in {\tt PYTHIA 8} and reconstructed with {\tt Delphes 3}. After matching, each bin has a number of events smaller than the one generated at parton level and an associated matched cross section, $\sigma_{\mathrm{LO}-\mathrm{matched}}$. These values must be incorporated into the event weights to give a proper relative normalization to the bins. More precisely, each event contributes to a certain histogram with a weight given by
$$\frac{K\sigma_{\mathrm{LO}-\mathrm{matched},\,i}}{N_{i}},$$
where $\sigma_{\mathrm{LO}-\mathrm{matched},\,i}$ and $N_i$ are the cross section and the number of events after matching obtained for the bin $i$ ($=1,..,6$ in our case), and $K$ is the $K$-factor applied to include QCD corrections. 

\begin{figure}[t!]
\begin{center}
\includegraphics[width=.49\textwidth]{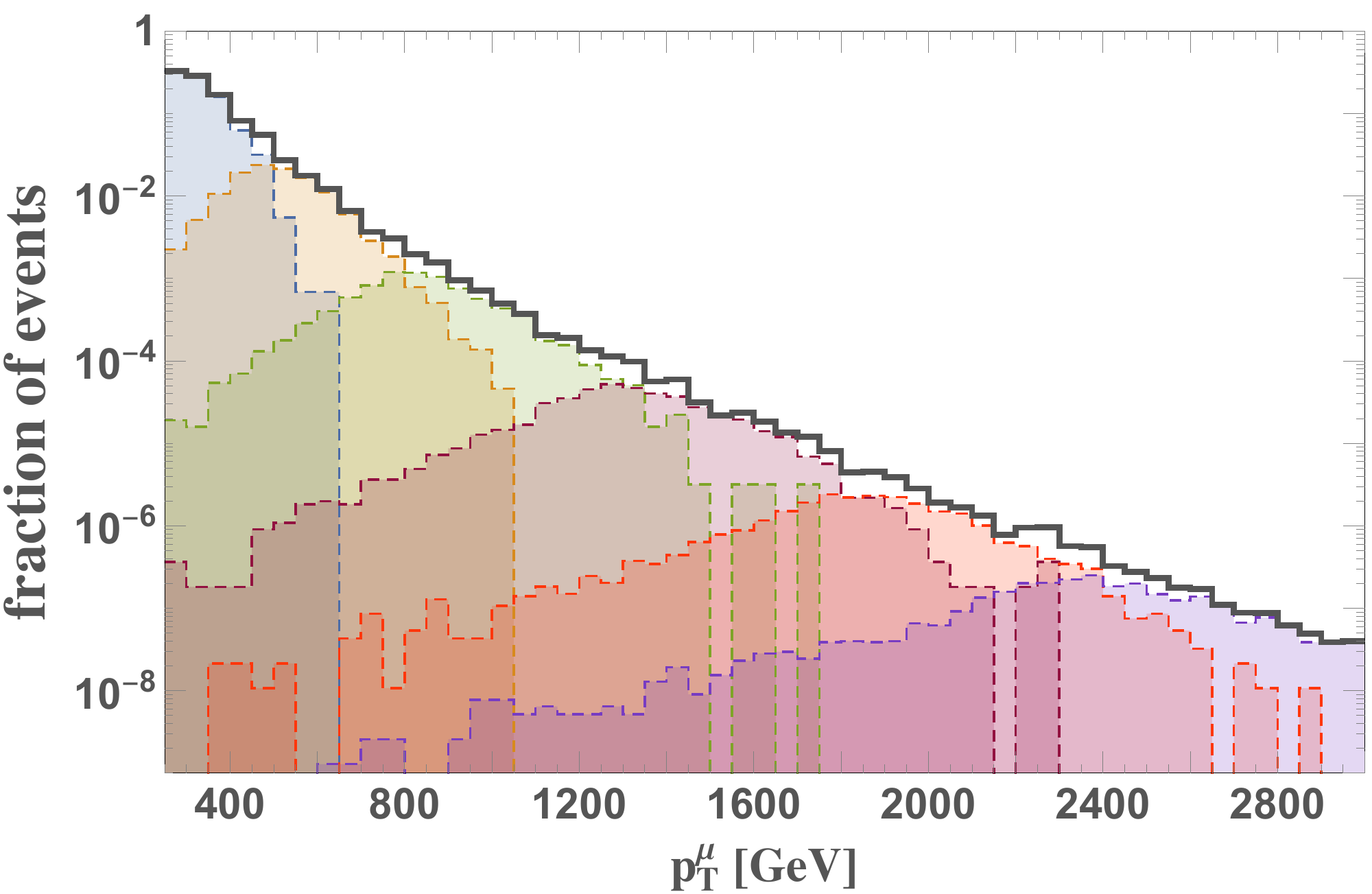} 
\includegraphics[width=.49\textwidth]{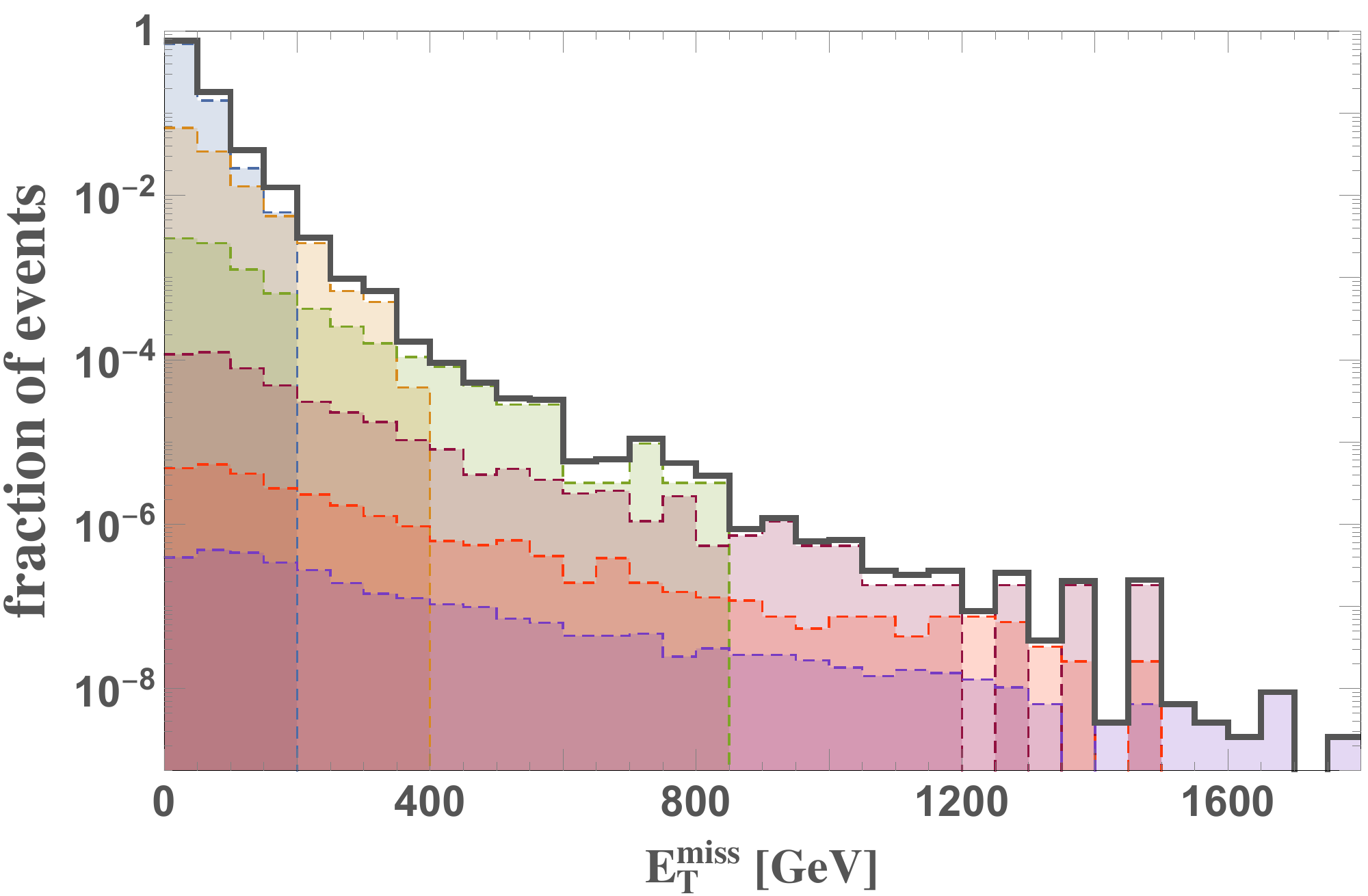} 
\caption{Distribution of $p_T^\mu$ and $E^{\mathrm{miss}}_T$ in multijet events after applying the requirements in Eq.~(\ref{basic-cuts}). We show the individual contributions of each of the six $H_{T2}$ bins (dashed lines) as well as the final distribution obtained by combining all of them (black solid line).}\label{METmultijet}
\end{center}
\end{figure}

With this normalization, the sum over all the events after matching of the weights gives the total cross section of the multijet process corrected at NLO with the $K$-factor. By using the sample of events generated in the six bins of $H_{T2}$ we construct the relevant kinematic distributions and determine the acceptances corresponding to the multijet background of the different cuts applied in our search strategy.\par 
As an illustration of the simulation procedure we show in Fig.~\ref{METmultijet} the individual contribution of each of the six bins of $H_{T2}$ to the $p_T^\mu$ and $E^{\mathrm{miss}}_T$ distributions, along with the histograms obtained by adding all of them. As can be seen from this figure the simulation in exclusive bins of $H_{T2}$ provides an accurate modeling of the distributions, specially of their tails, which is crucial to obtain a reliable estimation of the acceptances of lower cuts. Moreover, this is achieved with considerably less computational resources than in the case of the unweighted approach.

\bibliographystyle{unsrt}

\end{document}